# Understanding and optimizing the packing density of perylene bisimide layers on CVD-grown graphene

*Nina C. Berner [‡†], Sinéad Winters [‡†|], Claudia Backes[⊥], Chanyoung Yim[†|], Kim C. Dümbgen[†], Izabela Kaminska[₸], Sebastian Mackowski[₸], Attilio A. Cafolla[∩], Andreas Hirsch[Δ], and Georg S. Duesberg*[†|]*

[†]Centre for the Research on Adaptive Nanostructures and Nanodevices (CRANN) and Advanced Materials and BioEngineering Research (AMBER), Trinity College Dublin, Dublin 2, Ireland

[|]School of Chemistry, Trinity College Dublin, Dublin 2, Ireland

[⊥]School of Physics, Trinity College Dublin, Dublin 2, Ireland

[₸]Faculty of Physics, Astronomy and Informatics, Nicolaus Copernicus University, Grudziadzka 5, 87-100 Torun, Poland

[∩]School of Physical Sciences, Dublin City University, Dublin 13, Ireland

[Δ]Institute of Organic Chemistry II, University of Erlangen-Nürnberg, Henkestr. 42, 91054 Erlangen, Germany

**ABSTRACT:** The non-covalent functionalization of graphene is an attractive strategy to alter the surface chemistry of graphene without damaging its superior electrical and mechanical properties. Using the facile method of aqueous-phase functionalization on large-scale CVD-grown graphene, we investigated the formation of different packing densities in self-assembled monolayers (SAMs) of perylene bisimide derivatives and related this to the amount of substrate contamination. We were able to directly observe wet-chemically deposited SAMs in scanning tunneling microscopy (STM) on transferred CVD graphene and revealed that the densely packed perylene ad-layers adsorb with the conjugated π-system of the core perpendicular to the graphene substrate. This elucidation of the non-covalent functionalization of graphene has major implications on controlling its surface chemistry and opens new pathways for adaptable functionalization in ambient conditions and on the large scale.

**INTRODUCTION**

Since the isolation of single layer graphene by mechanical exfoliation[1] and the subsequent discovery and demonstration of its outstanding electronic[2] and mechanical[3] properties, graphene has attracted an extremely high level of interest. It has been proposed for numerous applications in electronics,[4] photonics,[5-6] sensing,[7] as well as gas barriers[8] and coatings.[9] However, for most of these applications it is necessary to introduce functional groups onto the graphene surface, *e.g.* to achieve sensor selectivity.

Non-covalent functionalization is an attractive strategy to introduce chemical functionalities since it does not adversely affect the electronic properties of the graphene layer.[10-12] Organic molecules with a relatively large and planar aromatic core have been repeatedly reported, both experimentally and theoretically, to adsorb on the graphene surface *via* van der Waals interactions between the extended π-orbital systems of the molecular core and the graphene, often referred to as π-π stacking.[13-15] Most detailed studies on the non-covalent functionalization of graphene have been conducted on either chemically exfoliated graphene flakes in solution[16] or on epitaxially grown graphene on SiC substrates in ultra-high vacuum (UHV).[17-18] However, neither of these graphene preparation methods is suitable for the fabrication of devices on an industrial scale. Graphene grown by chemical vapor deposition (CVD) is better suited for industrial electronic applications, especially since the growth and transfer onto $SiO_2$ substrates have been vastly improved in recent years[19-20] and CVD can now produce graphene of high structural integrity and with electrical characteristics that can compete with those of mechanically exfoliated graphene.[21-22]

Non-covalent functionalization of CVD graphene can be achieved by thermal evaporation[23-24] or by deposition of the molecules from liquid phase.[25-27] However, a few recent studies have shown that the mode of deposition[28] and the nature of the graphene substrate[29] have a significant influence on the adsorption geometry and molecular orientation in the ad-layers, which is a very important factor since it has been shown to influence the film's characteristics such as light adsorption, charge transport and energy level alignment[30]. These findings imply that many of the reported results obtained on *e.g.* epitaxially grown graphene and with molecules evaporated under UHV conditions may not be applicable to the same or very similar material systems obtained by scalable methods like CVD growth and subsequent wet-chemical functionalization. Nevertheless, investigating the thus obtained material systems is challenging for a number of reasons. It has been demonstrated that electrical measurements obtained from wet-chemically deposited molecule films on graphene can be difficult to reproduce due to unwanted and uncontrollable doping and disturbance of the structural uniformity of the molecular layers by solvent co-adsorption.[26-27] In addition, the presence of transfer polymer residue on CVD graphene has been shown to affect the organization of the molecular layers.[29, 31] These effects can be significantly reduced by optimizing the quality of the graphene surface. Here we describe the functionalization of graphene with a water-soluble perylene bisimide derivative **1**, shown in Figure 1a, from the liquid phase and its characterization with water contact angle measurements, Raman and fluorescence spectroscopy, spectroscopic ellipsometry (SE) as well as, for the first time at room temperature, STM with atomic resolution of an organic ad-layer directly on transferred CVD graphene in UHV.

Molecules similar to **1** have previously been used as surfactants in liquid exfoliation of carbon allotropes.[16, 32-33] It was assumed that the aromatic perylene core is adsorbed on the carbon allotrope *via* π-π stacking, while the carboxylic acid groups terminating the dendritic side chains stabilize the dispersed carbon allotropes in water. However, such molecular systems are also of great interest in terms of non-covalent functionalization of CVD graphene, as the carboxylic acid groups are highly potent anchor groups for further derivatization. It is therefore important to understand and tune the adsorption behavior. In this study, we demonstrate the influence of the cleanliness of the graphene surface on the packing density of the films and investigate the geometry of the most densely packed self-assembled monolayers.

**EXPERIMENTAL SECTION**

Raman studies were conducted in air, using a WiTec Alpha 300 nm confocal Raman system with a 532 nm excitation laser wavelength. Raman maps were acquired with a point spectrum taken every 400 nm and an integration time of 0.1 s/spectrum. The intensity of all spectra was normalized to the intensity of the graphene G peak before further processing, *i.e.* the calculation of intensity ratios and the mapping thereof. All analysis was done using the WiTec Project software.

Scanning Tunneling Microscopy (STM) measurements were performed in UHV (pressure < $10^{-10}$ mbar) at room temperature on an Omicron Variable Temperature (VT)-STM. Samples were attached to tantalum sample plates using tantalum wire, which also helped to ensure electrical contact between the graphene film and the sample plate. The STM tips used in these experiments were electrochemically etched from tungsten wire (0.15 mm diameter) using 2M KOH. The imaging parameters used for each image displayed in this manuscript are provided individually with every figure. The images underwent minimal processing steps limited to FFT-supported flattening and Gaussian blurs (over 2px) using the software WSxM.[34]

An Alpha SE tool (J. A. Woollam Co., Inc.) was used for the Spectroscopic Ellipsometry (SE) data measurements, operating in the wavelength range of 380 – 900 nm (1.38 – 3.25 eV) at an angle of incidence of 65° with a beam spot size of ~40 mm$^2$. The measured data was analyzed using the CompleteEASE software (Ver.4.72, J. A. Woollam Co., Inc.). The SE system gathered values of psi ($\Psi$) and delta ($\Delta$), which represent the amplitude ratio ($\Psi$) and phase difference ($\Delta$) between p- and s-polarizations, over the specific wavelength range. The two parameters are related to the ratio $\rho$, defined by the equation of $\rho = r_p/r_s = \tan(\Psi)\exp(i\Delta)$, where $r_p$ and $r_s$ are the amplitude reflection coefficients for the p-polarized and s-polarized light, respectively.

Spectrally- and time-resolved fluorescence measurements were performed using a home-built confocal fluorescence microscope described in detail elsewhere.[35] The sample was placed on a piezoelectric translation stage, which enabled continuous movement of the sample with respect to the excitation laser beam. We used pulsed laser excitation at 485 nm (with a repetition rate of 20 MHz and power of 10 µW). Fluorescence was extracted using a combination of FEL550 + 665/40 filters.

Wide-field microscopy imaging experiments of the perylene on graphene samples were carried out using an inverted fluorescence wide-field Nikon Eclipse Ti-U microscope equipped with an Andor iXon Du-888 EMCCD detector. Immersion objective with magnification 100x (Plan Apo, Nikon) and the numerical aperture 1.4 was used, which provides spatial resolution of about 300 nm. As a light source we used an LED illuminator (485 nm) equipped with a band pass filter (FB 480-10). Excitation power was equal to 100 µW. Fluorescence of **1** on graphene was extracted by combining a dichroic mirror (Chroma 505dcxr) and a band pass filter (Thorlabs FB 655-40).

Graphene was grown on copper foil by CVD, using methane as the carbon source at a temperature of 1035°C, as described in more detail previously.[19] Sample-sized pieces (typically around 1 x 1 $cm^2$) were subsequently transferred on a $Si/SiO_2$ (300 nm or 150 nm) substrate using the established PMMA-assisted method. In a typical process, a 100 nm film of PMMA was spun-cast on the graphene grown on the copper substrate. The copper was subsequently etched in a 1M ammonium persulfate solution. The remaining PMMA/graphene layer was floated onto de-ionized water and left to rinse for 1 hour. The film was then transferred to the substrates and left to dry in air. The PMMA was removed by immersion in acetone overnight. To produce annealed samples with reduced PMMA residue, substrates were annealed overnight in UHV (pressure < $10^{-9}$ mbar) at a temperature of 220°C.

Perylene **1** was synthesized as described elsewhere[36] and then dissolved in aqueous sodium phosphate buffer solution (pH 7, 0.1M) at a concentration of 0.001 mol $L^{-1}$. The solution was applied to the graphene surface by drop-casting or dip-casting for 5-10 seconds and subsequent rinsing with de-ionized water and isopropyl alcohol. The samples were blow-dried with dry nitrogen before characterization. Different perylene concentrations and deposition times were found to have negligible effects on the layer formation, as briefly discussed in the SI.

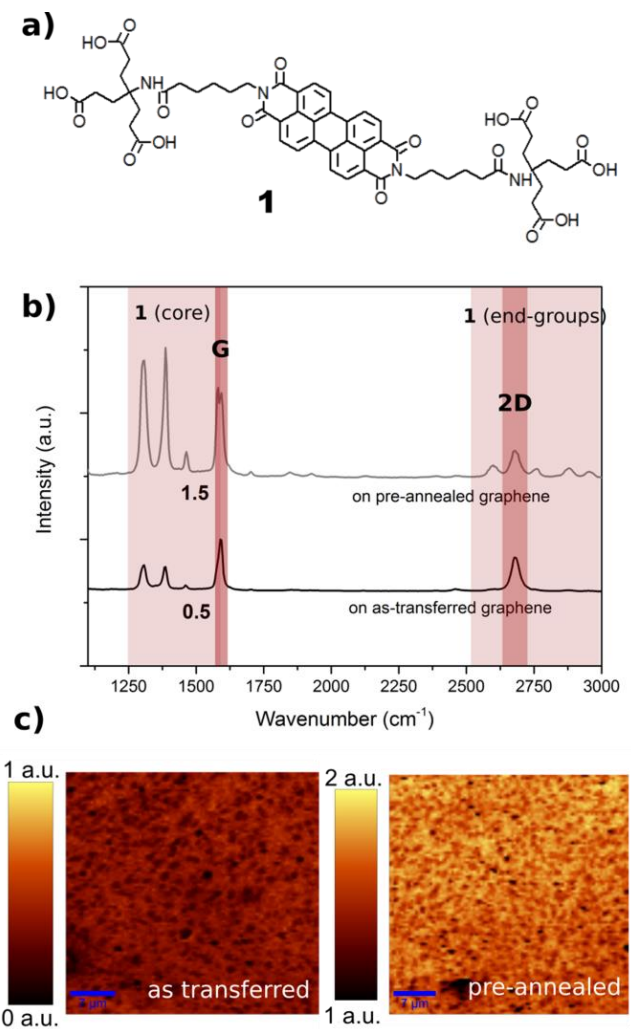

**Figure 1.** a) Chemical structure of Perylene **1**, b) average Raman spectra of **1** deposited on as-transferred and annealed CVD-grown graphene on $SiO_2$, with peak assignment labels for colored wavenumber regions, c) corresponding Raman maps of **1** (core):G peak ratio in LPD and HPD layers.

**RESULTS AND DISCUSSION**

After the deposition of **1** onto CVD-grown graphene from aqueous solution, as described in detail in the Methods section, the layers were investigated with Raman spectroscopy. The perylene molecules can be easily identified and characterized by their two major characteristic Raman peaks at 1303 cm$^{-1}$ and 1383 cm$^{-1}$ when resonantly excited at 532 nm, as previously described and demonstrated by Kozhemyakina *et al.* on liquid-exfoliated graphene.[37] The ratio of those two most intense **1** peaks to the G peak of the graphene substrate (**1**:G) can be seen as an indicator of the packing density of the molecular layer. The black curve in Figure 1b shows a Raman spectrum of **1** deposited on CVD-grown graphene after it was transferred onto $SiO_2$ using a standard PMMA-assisted process as described in detail elsewhere.[19] Figure 1c shows the corresponding Raman intensity maps of the **1**:G intensity ratio over a 30 x 30 μm$^2$ area, showing very high uniformity except in small patches which correspond to 2nd layer graphene growth. The **1**:G peak ratio was determined to be 0.5 on average. With very high reproducibility, the packing densities of the molecular films are much higher when **1** was deposited on a transferred graphene substrate which had been previously annealed to reduce PMMA residue,[38] as can be seen in the grey spectrum in Fig. 1b and the **1**:G ratio map in Figure 1c. We attribute this to the absence of polymer or other hydrocarbon contaminants on the pre-annealed graphene surface, which allows the perylene molecules to rearrange more easily and to self-assemble into a densely packed layer. Further details of additional peaks in the Raman spectrum of **1** are provided in the SI.

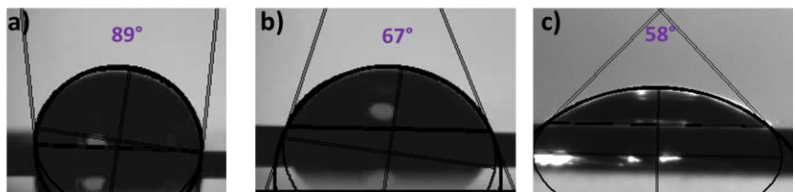

**Figure 2.** Water contact angle measurements of a) clean graphene, b) low packing density **1** layer on graphene, and c) high packing density **1** layer on graphene.

The different packing densities of **1** on graphene also have an effect on the layer thickness as estimated by spectroscopic ellipsometry (SE). From a model that is discussed in detail in the SI, the **1** layers on as-transferred and pre-annealed graphene substrates were measured to be 2.2 (± 0.1) nm and 5.4 (± 0.2) nm thick, corresponding to the low (LPD) and high packing density (HPD), respectively. The difference between these values underlines the significance of the packing density variation.

The significantly increased packing density of **1** on cleaner graphene is further confirmed by water contact angle (WCA) measurements, typical examples of which are depicted in Figure 2. Bare graphene samples on $SiO_2$ have a water contact angle of 89° (±1°), similar to what has been reported elsewhere.[39] The contact angle is decreased upon addition of **1** to the surface, due to the presence of hydrophilic groups, in particular the six carboxylic acid functions. For graphene samples with LPD perylene molecules, a water contact angle of 67° (±2°) is measured, while an angle of 58° (±2°) is observed at HPD layers of **1**, clearly indicating an increase in the number of carboxylic acid groups on the surface.

The exact nature of the HPD layers is further investigated by characterization of the transferred graphene substrate and subsequent perylene **1** deposition by scanning tunneling microscopy (STM). Figure 3a shows an STM image of CVD-grown graphene transferred with PMMA on SiO$_2$ before and after an overnight anneal in UHV at 220°C. As clearly visible in the upper part of Figure 3a, images taken on the as-transferred surface show streaky features that indicate loosely bound material which is moved over the surface by the STM tip. In contrast, images taken on the same surface after the overnight anneal are much clearer and show almost no evidence of adsorbates. After deposition of **1** on the pre-annealed, clean graphene substrate and a short anneal to just over 100°C to remove any adsorbed water, STM images taken over the entire surface area show periodic structures as shown in Figure 3b, which, upon closer inspection, can be related to molecular arrays. Figure 3c shows line profiles across some of the features which reveal periodicities of 0.23 nm and 3.25 nm along orthogonal directions (further images and analysis are shown in the SI). This leads us to the conclusion that in its HPD configuration, **1** adsorbs and packs into SAMs with the perylene cores perpendicular to the surface, as indicated by the molecular structure overlay in Figure 3d and visualized in Figure S4e and the TOC graphic. This observation is in contrast to the common assumption that organic molecules with large conjugated π-systems adsorb via π-π interactions between the molecular cores and the graphene substrate, regardless of the method of deposition and preparation method of the graphene. Our findings emphasize that the adsorption of molecules on surfaces is governed by a complex interplay of interfacial energies between both the molecules themselves and the molecules and the substrate's surface. If molecules are deposited from solution, they are surrounded by a much larger number of neighboring molecules than if they were deposited by thermal evaporation, which changes the energy landscape and therefore the adsorption behaviour significantly. This is particularly the case for **1** which is not present as monomer in solution, but forms micelles due to the strong interaction between the perylene cores and the amphiphilic nature of the molecules.[36] The other major influencing factor is the surface energy of the substrate, which is evidenced by the difference in adsorption behaviour of **1** when the graphene substrate is contaminated with PMMA residue. In the particular case of the HPD layer of **1** on clean CVD graphene, it appears that the molecule-molecule interactions are much stronger than the molecule-substrate interactions, but this may be different in an even slightly altered environment, for example by residual PMMA on the substrate. This is indirectly supported by STM images recorded on samples with LPD of **1**, an example of which is shown in Figure S4f. None of the images showed any clearly resolved features, which indicates an abundance of loose material on the surface and the lack of molecule-molecule stabilization in organized structures. Therefore the adsorption geometry may be different for LPD perylene films. In fact, the lower density of carboxylic acid groups evidenced by the WCA measurements implies that at least some of the molecules are adsorbed with the perylene core lying flat on the substrate.

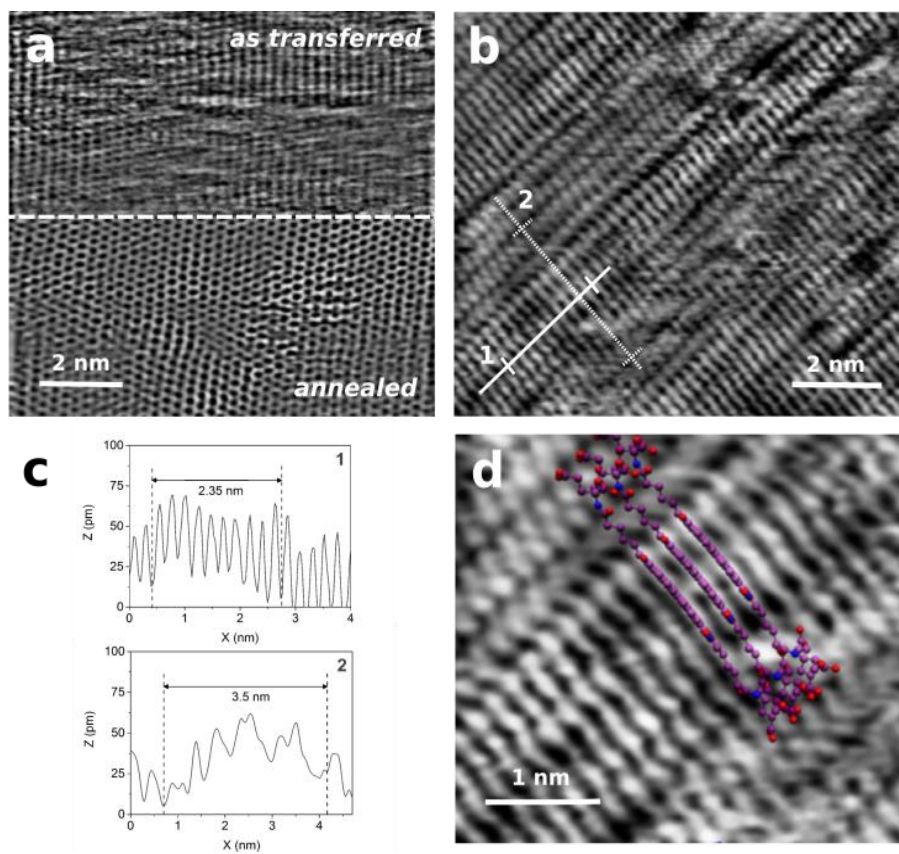

**Figure 3.** STM images of a) as-transferred (top, taken at -0.3 V/0.3 nA) and subsequently annealed (bottom, taken at -0.5 V/0.8 nA) CVD graphene on $SiO_2$ and b) after wet-chemical deposition of **1** on the same annealed substrate (taken at -0.5 V/0.1 nA); c) line profiles from b) and zoomed in area with molecular structure overlay (d).

However, the results of fluorescence imaging and spectroscopy also provide some evidence of vertical packing of **1** on graphene in the LPD structure. Figure 4a shows an example of a fluorescence image obtained for a 100 x 90 μm$^2$ area. While the intensity is rather low and some completely dark areas are visible, a large fraction of the sample exhibits measurable emission. The overall low emission intensity is a result of both monolayer coverage of **1** on the graphene substrate and energy transfer to graphene. Spectrally-resolved and time-resolved data (Figures 4b and c), obtained for fluorescent regions on the sample shows that the emission of **1** appears around 670 nm, which is dramatically shifted to longer wavelengths as compared to the solution.[40] This points towards a strong interaction between the aromatic rings of the molecules, which is possible only in a configuration where the molecules are placed at an angle to the substrate. Furthermore, a much shorter fluorescence decay evidences an energy transfer with a low efficiency of less than 80%, which again suggests that not all of the molecules are lying flat on the surface, as this would promote essentially complete quenching of the fluorescence.[41] The observations obtained for the LPD structure are corroborated with the results of analogous experiments for regions with multilayer coverage, where several-layer-thick regions are present in addition to monolayer areas (see SI). In particular, the spectral position of high-intensity regions is exactly the same as for monolayer regions, which further supports that the molecular orientation vertical to the graphene substrate is the most energetically favorable and that molecule-molecule interactions within the same self-assembled layer are the dominating factor, even in multilayers.

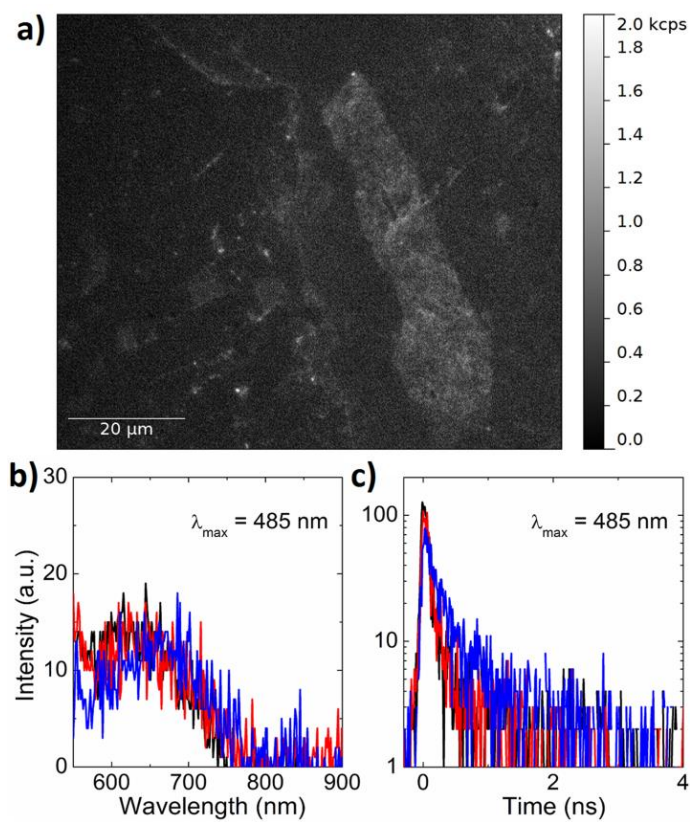

**Figure 4.** a) Wide-field fluorescence image of LPD **1** on graphene. b) Typical fluorescence spectra and c) fluorescence decay curves obtained for three different locations across the LPD sample. Excitation at 485 nm was used in both experiments.

**CONCLUSION**

In summary, we report on the non-covalent functionalization of CVD-grown and transferred graphene from aqueous solution. We demonstrated the dependency of the packing density of perylene bisimide derivative ad-layers on the surface contamination of the graphene substrate, as easily recognizable from Raman spectroscopy, SE and WCA measurements. Furthermore, fluorescence spectroscopy and STM images of densely packed perylene films on CVD graphene reveal a vertical adsorption geometry stabilized by π-π interactions between the cores of the molecules. The altered surface energy of the graphene caused by transfer polymer residue appears to alter the adsorption geometry of the perylenes and cause them to adsorb with the core flat or at least at a lower angle to the substrate in some regions, while still standing upright in others. This increased understanding of the adsorption and self-assembly of wet-chemically deposited organic molecules on CVD-grown and transferred graphene highlights the often underestimated complexity of molecular adsorption on graphene and is therefore an important step towards the reliable large-scale fabrication of non-covalently functionalized 2D materials and their application.

**ACKNOWLEDGMENT**

NCB, SW and GSD acknowledge the SFI under grant number PI_10/IN.1/I3030. The research leading to these results has also received funding from the European Union Seventh Framework Program under grant agreement n°604391 Graphene Flagship. In addition, CB acknowledges the German research foundation DFG (BA 4856/1-1). IK and SM are supported by the project number DEC-2013/10/E/ST3/00034 from the National Science Center (NCN). AH would like to thank the DFG (SFB 953 Synthetic Carbon Allotropes) for financial support. AAC thanks the SFI, grant number 08/RFP/PHY1366. We thank C. Wirtz for providing the schematics.

**ASSOCIATED CONTENT**

**Supporting Information Available**. Additional Raman spectra of the graphene substrates before and after annealing and deposition of **1**, fluorescence images of multilayer regions of **1**, a more detailed discussion of SE results, and some additional STM images can be found in the Supporting Information (SI). This material is available free of charge via the Internet at http://pubs.acs.org.

**AUTHOR INFORMATION**

**Corresponding Author**

*Georg S. Duesberg, e-mail: duesberg@tcd.ie.

**Author Contributions**

The manuscript was written through contributions of all authors. All authors have given approval to the final version of the manuscript. ‡These authors contributed equally.

**TOC GRAPHIC**

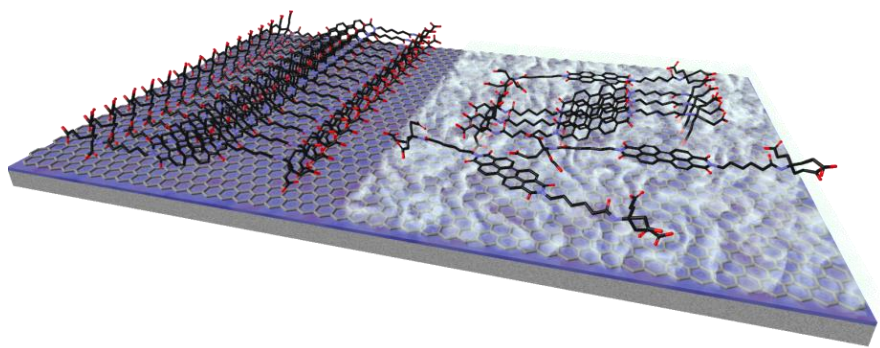

# SUPPORTING INFORMATION

Understanding and optimizing the packing density of perylene bisimide layers on CVD-grown graphene

*Nina C. Berner [‡†], Sinéad Winters [‡†∥], Claudia Backes[⊥], Chanyoung Yim[†∥], Kim C. Dümbgen[†], Izabela Kaminska[Ŧ], Sebastian Mackowski[Ŧ], Attilio A. Cafolla[∩], Andreas Hirsch[Δ], and Georg S. Duesberg*[†∥]*

[†]Centre for the Research on Adaptive Nanostructures and Nanodevices (CRANN) and Advanced Materials and BioEngineering Research (AMBER), Trinity College Dublin, Dublin 2, Ireland

[∥]School of Chemistry, Trinity College Dublin, Dublin 2, Ireland

[⊥]School of Physics, Trinity College Dublin, Dublin 2, Ireland

[Ŧ]Faculty of Physics, Astronomy and Informatics, Nicolaus Copernicus University, Grudziadzka 5, 87-100 Torun, Poland

[∩]School of Physical Sciences, Dublin City University, Dublin 13, Ireland

[Δ]Institute of Organic Chemistry II, University of Erlangen-Nürnberg, Henkestr. 42, 91054 Erlangen, Germany

**Graphene substrates – Raman spectroscopy**

Representative Raman spectra of graphene after transfer to a $SiO_2$ substrate using the process described in the Methods section of the main manuscript and after subsequent annealing in UHV are shown in Fig. S1a. The 2D peak can be fitted with a single Lorentzian function (with a FWHM of 27 cm$^{-1}$) and the G:2D ratio is 1.75, both of which is characteristic for a single layer of high-quality graphene.[1] Additionally, the D peak at 1343 cm$^{-1}$ is comparatively small and indicates a low amount of defects, confirming the high quality of the CVD graphene substrates. Virtually no change was observed in the spectrum after the annealing step to remove PMMA residue, which confirms the preservation of the good graphene quality.

Graphene intended for STM measurements was transferred onto Si/SiO$_2$ wafer pieces which were significantly smaller than the graphene film and draped over the edges to make contact with the conducting STM sample plate.

Figure S1b shows a large scale (300 nm x 300 nm) STM image taken of CVD graphene transferred onto 150 nm SiO$_2$ immediately after introduction into the vacuum and a brief anneal to just over 100°C to remove any adsorbed water. It shows some graphene wrinkles, as previously observed using AFM,[2] in the top left corner, and the graphene film following the irregular surface structure of the SiO$_2$ substrate in the rest of the image. The maximum observed height of graphene wrinkles was ~3 nm, whereas the roughness of the "flat" graphene/SiO$_2$ parts of the sample was max. 0.5 nm over the scanned area. In order to increase the contrast of smaller features like the honeycomb structure of graphene, the irregular underlying SiO$_2$ structure was removed by FFT-aided flattening in all high-resolution images as best as possible. Figure S4a and b show an STM image after and before image manipulation, respectively, demonstrating this process.

**Perylene films on graphene – additional Raman spectra**

Raman spectra of **1** on graphene show many features in addition to the characteristic G and 2D bands of the CVD graphene spectrum, as can be seen in Figures S2b and c. The two most distinct peaks characteristic for perylene compounds and originating from their core sit at 1303 cm$^{-1}$ and 1383 cm$^{-1}$ and can be conveniently used for mapping molecular coverage, as demonstrated in Fig. S2a. Additionally, as described in the main manuscript, the intensity ratio of these two characteristic Raman peaks of **1** and the normalized G peak at 1591($\pm$2) cm$^{-1}$ can be used as a measure of the packing density in the molecular ad-layer. Some additional minor peaks associated with the head groups of the dendrimers of **1** (carbonyl/caryboxyl stretching vibrations) can be observed around the 2D peak of graphene (as shown and labeled in Figure 1b in the main manuscript) in HPD layers, but are not visible in LPD layers due to their comparatively low intensity. Figure S2a shows a map of the **1** Raman features over an area which includes both monolayer graphene as well as the bare SiO$_2$ substrate and folded over graphene, as can be seen in the corresponding optical image. Some points are highlighted in the map and the corresponding point spectra are shown in Fig. S2b. The Raman signature of **1** is higher in areas where the graphene/**1** layer is folded over (blue spectrum in Fig. S2b), as can be expected due to the higher density of molecules, and it is not visible at all on the SiO$_2$ (black spectrum), which can most likely be attributed to the lack of the enhancement effect of graphene on the Raman signal of organic molecules.[3] It should be noted that this rather defective area of a sample was chosen for display and discussion due to its many interesting features, and that the quality of the graphene/**1** films is usually much higher, without defects and with uniform coverage, as indicated in Figure 1c in the main manuscript.

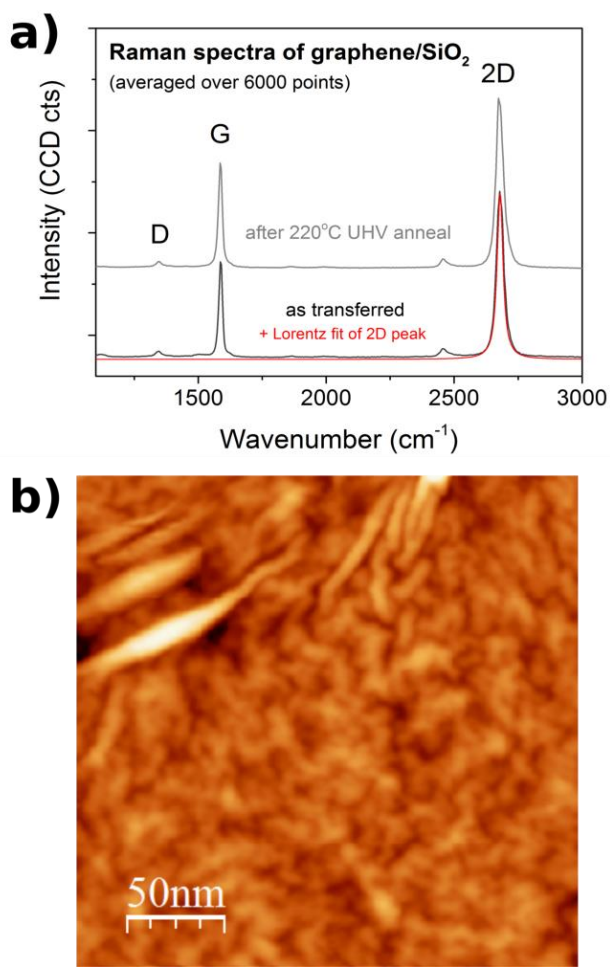

**Figure S1.** a) Raman spectra of a CVD-grown graphene film transferred onto SiO$_2$ before and after annealing, averaged over an area of 30 μm x 30 μm. b) STM image of the same graphene sample on SiO$_2$, taken at -1.8 V/1 nA.

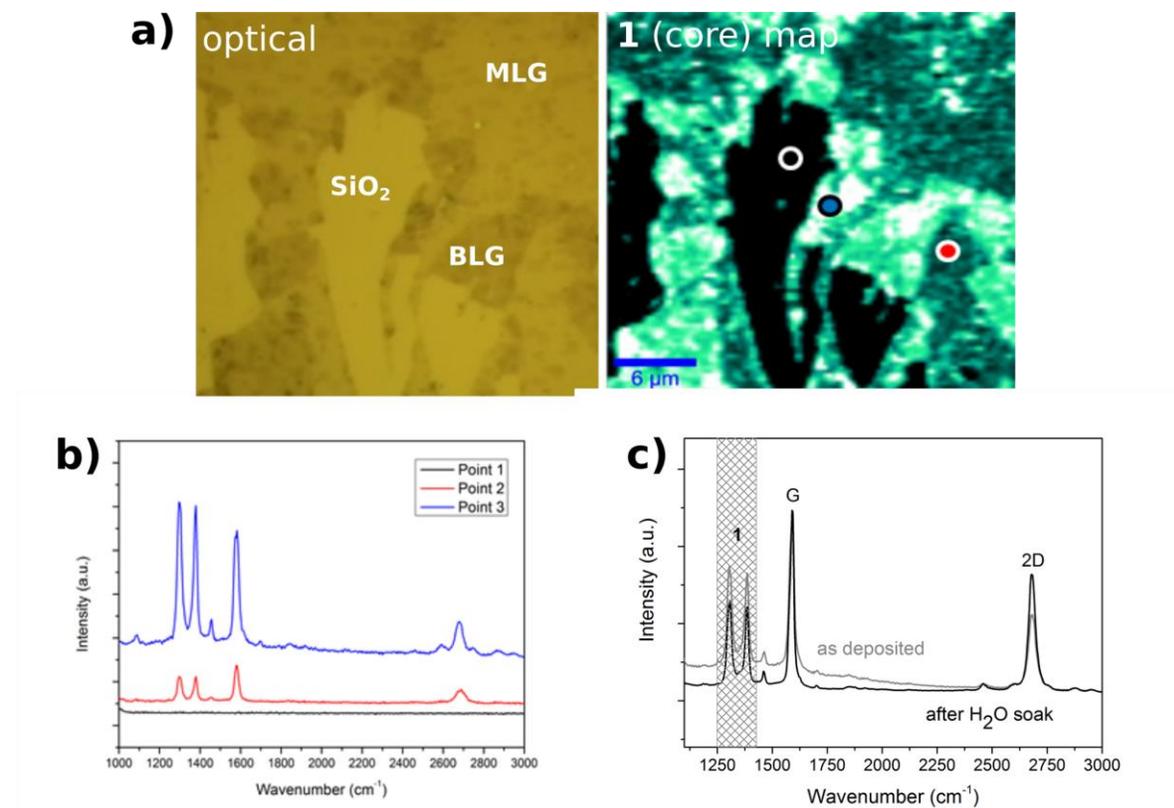

**Figure S2.** a) Optical image of an area of monolayer graphene/**1** (MLG/**1**) with areas of exposed SiO$_2$ substrate and folded over graphene/**1** (BLG/**1**) and corresponding Raman map of the characteristic peaks of **1** (see spectra below, marked area in (c)). b) Raman spectra measured at the points marked in the map in (a). c) Large area averaged (6000 pts over 30 x 30 μm$^2$) Raman spectra of **1** on a different graphene sample directly after deposition from aqueous solution (grey) and after soaking in H$_2$O for 10 minutes (black).

It is furthermore important to discuss that **1** has another Raman peak in the immediate vicinity to the graphene G band. It is shifted by approximately -10 cm$^{-1}$ with respect to the G peak and is visible as a shoulder for lower packing densities (*i.e.* intensities), as can be seen in the lower spectrum in Figure 1b (in the main manuscript), and as a distinguishable peak at higher packing densities. This feature must not be confused with the G band or G band splitting due to strain in the graphene.[4]

Raman spectra of **1** directly after deposition onto graphene often show a small fluorescence background and fluctuations in packing density. We attribute this to the formation of small areas of a 2nd layer or multilayers of **1**, since the inherent fluorescence of these molecules[5] is only quenched in the immediate vicinity of the graphene substrate,[6] but can be detected with the Raman spectrometer in multilayers as increased background intensity. As shown in Figure S2c, this background and therefore multilayer formation can be significantly reduced by soaking the samples in de-ionized water for at least 5 minutes.

**Perylene films on graphene – Concentration and deposition time**

The standard concentration of **1** in the aqueous buffer solution was chosen to be 0.001 mol L$^{-1}$, which was recommended by Backes *et al.* on the basis of their studies using **1** as a surfactant in graphene and SWCNT solutions.[7] To determine if the concentration has an impact on the adsorption characteristics, we varied it by a factor of 10 in both directions and found no difference in the Raman spectra of the thus formed molecular layers on graphene (not shown).

Additionally, we conducted a similar experiment varying the deposition time between 1 second and 20 minutes and again found no significant difference except a slight tendency towards increased multilayer formation (as evidenced by a small fluorescence background in the Raman spectrum, see Figure S2c for an example).

**Perylene films on graphene – Details on SE measurements**

As presented in Figure S3a, a six-layer optical model consisting of a Si substrate, an interface layer between Si and $SiO_2$, a $SiO_2$ layer, a graphene layer, another interface layer between graphene and perylene, and a perylene layer was built to interpret the SE spectra. A Cauchy model was used to extract the thickness of the perylene layer. The interface layer between graphene and perylene was built using an effective medium approximation (EMA) model,[8] which is composed of perylene and PMMA, considering the possible existence of PMMA residue on the graphene surface.

The perylene layer thickness was determined by fitting the experimental Ψ and Δ data with the simulated data from the optical model using a linear regression procedure. The fitting results of the Ψ and Δ data are plotted in Figure S3b and c, showing a good match between the experimental and simulated data. As mentioned in the main manuscript, the extracted thicknesses of the perylene molecule layers from this fitting procedure are 2.2 ± 0.1 nm for perylene on as-transferred graphene and 5.4 ± 0.2 nm for perylene on annealed graphene, respectively. Furthermore, the thicknesses of the interface layer between the graphene and perylene were found to be 0.6 ± 0.1 nm for perylene on as-transferred graphene and 0.2 ± 0.1 nm for perylene on annealed graphene, respectively, implying the surface roughness of the as-transferred graphene is higher than the annealed graphene most likely due to the effect of the PMMA residue on the graphene.

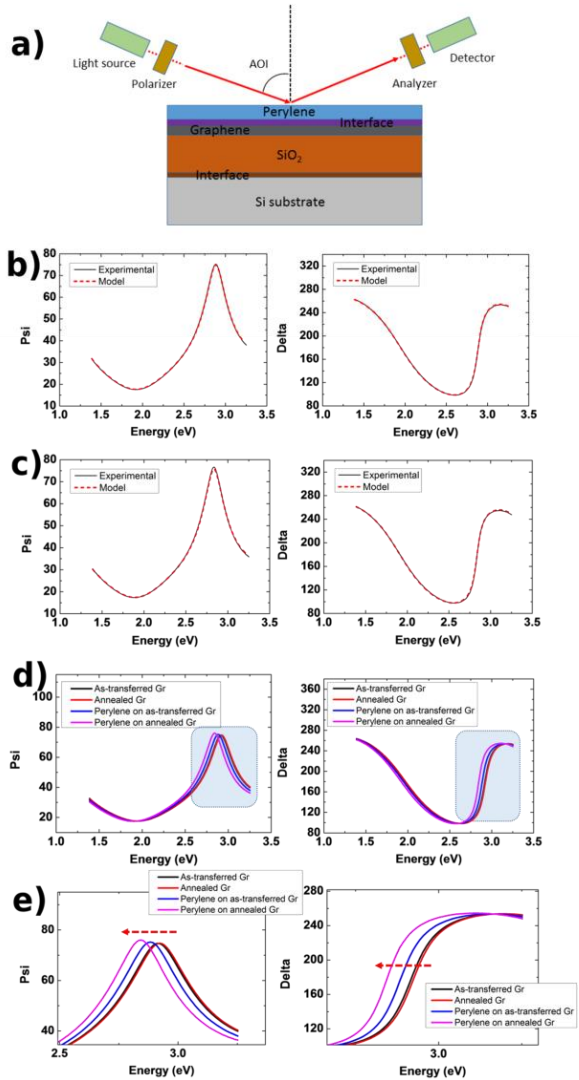

**Figure S3.** a) Schematic diagram of the optical model structure for SE data analysis. Fitting results of the psi (left) and delta (right) between experimental and simulated (model) data for the sample of (b) perylene on as-transferred Gr and (c) perylene on annealed Gr. (d) shows summary plots of psi and delta measured from four different samples (as-transferred Gr, annealed Gr, perylene on as-transferred Gr and perylene on annealed Gr) and (e) shows enlarged regions marked in (d).

In addition, comparing the measured SE spectra of four different samples which are as-transferred graphene (Gr), annealed Gr, perylene on as-transferred Gr and perylene on annealed Gr, clear peak shifts of the spectra depending on the top layer thickness change are observed in the range of 2.5 – 3.2 eV, which indicates this spectral region has a high sensitivity to the layer thickness variation of the samples.

**Perylene films on graphene – Additional STM images and analysis**

As discussed in the main manuscript, STM images of high packing density SAMs of **1** could be obtained after wet-chemical deposition on a clean and pre-annealed CVD graphene film on SiO$_2$. Figure S4a shows an additional STM image over a large undisturbed area (20 nm x 20 nm) of the observed adsorption pattern of **1**. Figure S4b shows the same image before any FFT-aided background flattening or other image manipulation was applied. Other areas on the sample showed the same pattern in the same orientation, no other domains have been found. A schematic rendering of the proposed structure of the SAM of HPD **1** on graphene as observed by STM is shown in Figure S4e.

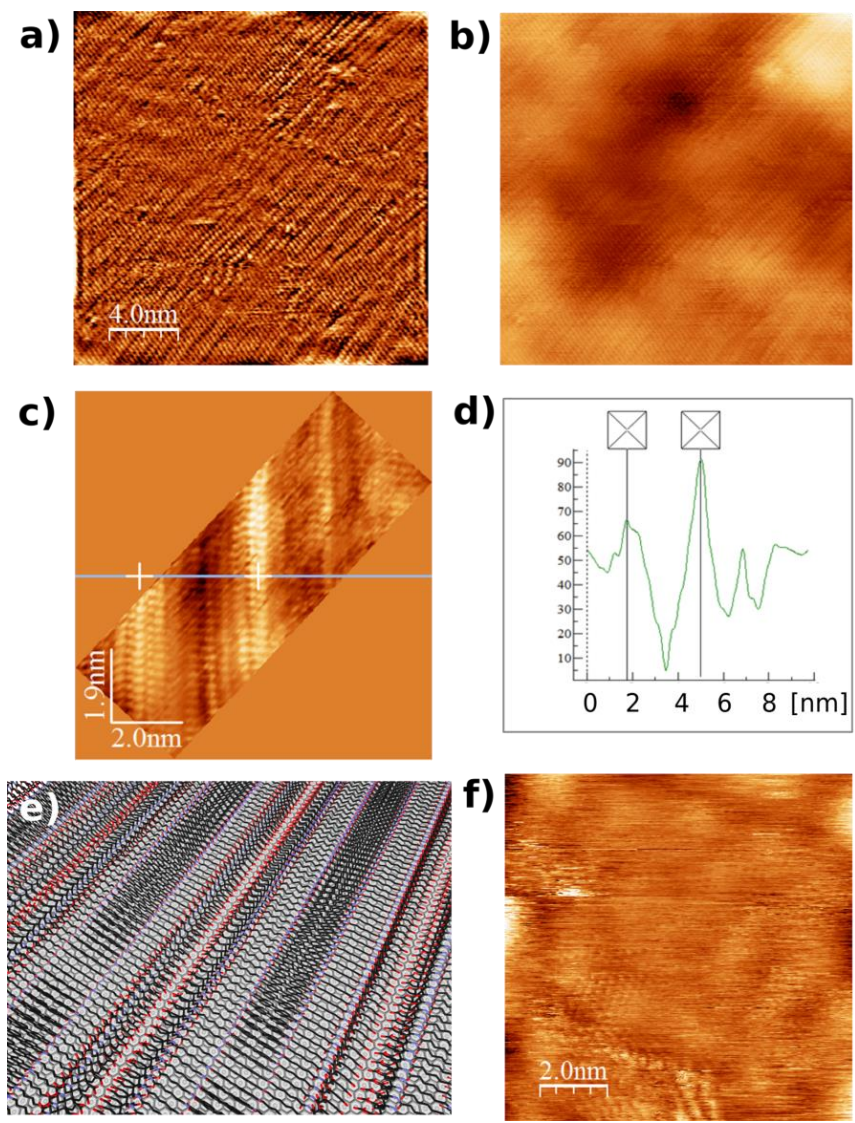

**Figure S4.** a) STM image of **1** on graphene/SiO$_2$ showing a large area (20 nm x 20 nm) without significant contamination or defects in the SAM; taken at -0.5 V/0.1 nA. b) The same image before processing and FFT-aided background flattening. c) High resolution STM image of **1** on graphene/SiO$_2$, taken at a lower bias (-0.3 V/0.2 nA), and rotated to align the **1** orientation with the x- and y-axes; d) profile of the y-integrated charge density along the x-axis with measurement of periodicity (i.e. molecule dimensions). e) Schematic of the high density SAM of **1** on graphene. f) Typical STM image of low density layers of **1** on as-transferred graphene/SiO$_2$, with streaky features indicating an abundance of loose material on the surface.

Figure S4c shows another high resolution STM image of **1** on graphene/$SiO_2$, taken at a lower bias than the other displayed ones, and rotated to align the **1** orientation with the x- and y-axes. A profile of the y-integrated charge density was taken along the x-axis and is shown in Figure S4d, including a measurement of the periodicity of the structure, *i.e.* the molecule dimensions.

Figure S4f displays a typical image obtained in the STM when attempting to image a **1** layer with low packing density, as signified by the respective Raman spectrum (not shown). As briefly mentioned in the main manuscript, the streaky features indicate loose material on the surface and prevent imaging with atomic resolution. This particular image, taken after repeated scanning of the same area, shows some areas with periodic features that could be related to the graphene substrate, but it is impossible to make any significant conclusions regarding the adsorption geometry of **1** in the low packing density layers from this or any other of the STM images taken on this surface.

**Multilayers of perylenes on graphene – Fluorescence imaging and spectroscopy**

Additional fluorescence images of areas with multilayer coverage of **1** on graphene are shown in Figure S5. Figure S6 displays the corresponding fluorescence spectra at different locations on the sample as well as the corresponding lifetime measurements. The latter indicate a correlation between high emission intensity and multi-exponential decay, including a long decay time component. In contrast, for low emission intensity, the decay is essentially identical to the one measured for the low packing density sample.

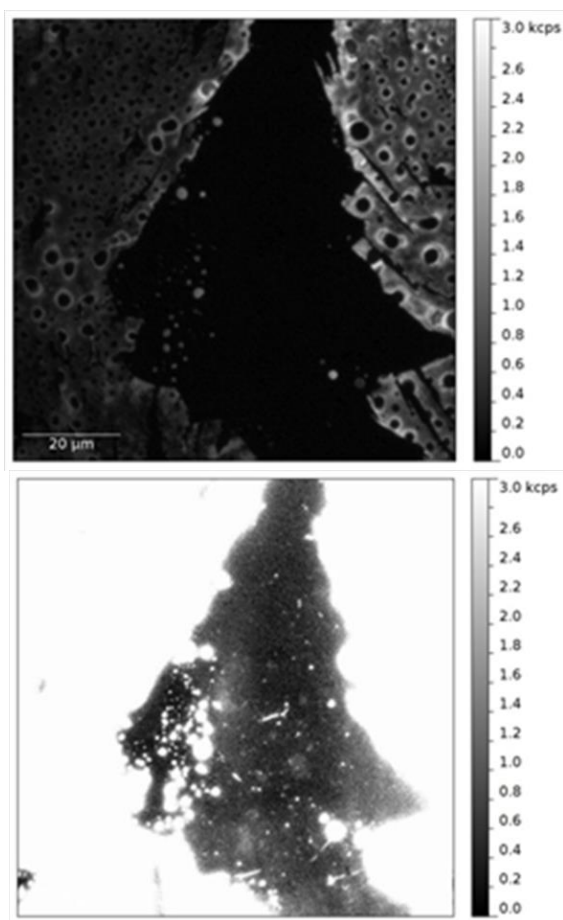

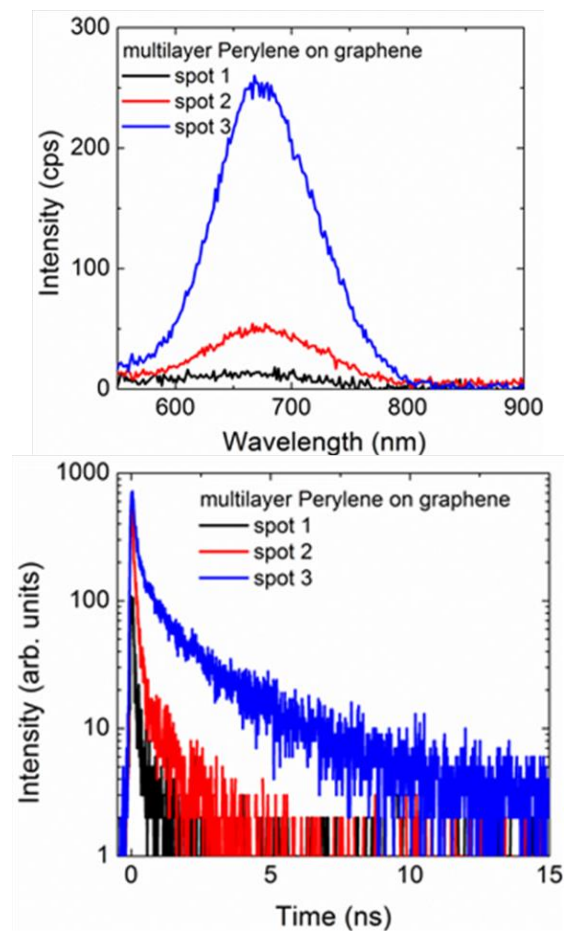

**Figure S5.** Fluorescence images obtained for samples with areas of multilayer coverage of **1** for two values of electron multiplying gain: 1 (upper), and 100 (lower). The lower map was taken in order to expose low-intensity areas similar to the ones visible in the image of the LPD layer in Figure 4 of the main manuscript.

**Figure S6.** Upper: Fluorescence spectra obtained for a sample with multiplayer coverage of **1** on graphene for three different locations across the sample, characterized with different emission intensities. The corresponding time traces shown in the lower panel.